\documentclass[12pt]{article}
\usepackage{latexsym}
\usepackage{epsfig}

\usepackage{graphicx}
\usepackage{epstopdf}

\usepackage{epsfig,amssymb,amsmath,amscd}

\newcommand{\bq}{\begin{equation}}
\newcommand{\eq}{\end{equation}}
\newcommand{\bqa}{\begin{eqnarray}}
\newcommand{\eqa}{\end{eqnarray}}
\newcommand{\ben}{\begin{enumerate}}
\newcommand{\een}{\end{enumerate}}
\newcommand{\bc}{\begin{center}}
\newcommand{\ec}{\end{center}}
\newcommand{\bqb}{\begin{eqnarray*}}
\newcommand{\eqb}{\end{eqnarray*}}

\def\pr#1#2#3{Phys. Rev. ${\bf{#1}}$, #2 (#3)}
\def\prl#1#2#3{Phys. Rev. Lett. ${\bf{#1}}$, #2 (#3)}
\def\pl#1#2#3{Phys. Lett. ${\bf{#1}}$, #2 (#3)}

\def\np#1#2#3{Nucl. Phys. ${\bf{#1}}$, #2 (#3)}

\def\zp#1#2#3{Z. f. Phys. ${\bf{#1}}$, #2 (#3)}

\def\epj#1#2#3{Eur. Phys. J. ${\bf{#1}}$, #2 (#3)}

\def\jmp#1#2#3{J. Mod. Phys. ${\bf{#1}}$, #2 (#3)}

\begin{document}
\pagenumbering{arabic}
\thispagestyle{empty}
\def\thefootnote{\fnsymbol{footnote}}
\setcounter{footnote}{1}

\begin{flushright}
March 14, 2017\\
 \end{flushright}

\begin{center}
{\Large {\bf Analysis of $\gamma\gamma\to ZH$ within the CSM concept}}.\\
 \vspace{1cm}
{\large F.M. Renard}\\
\vspace{0.2cm}
Laboratoire Univers et Particules de Montpellier,
UMR 5299\\
Universit\'{e} Montpellier II, Place Eug\`{e}ne Bataillon CC072\\
 F-34095 Montpellier Cedex 5, France.\\
\end{center}

\vspace*{1.cm}
\begin{center}
{\bf Abstract}
\end{center}

We study the modifications of the $\gamma\gamma\to ZH$ amplitudes
and cross sections generated by Higgs and top quark compositeness
in particular within the CSM concept. We insist on the particular
interest of polarized photon-photon collisions which should allow
to identify the origin of the large observable differences
between various, CSM conserving or CSM violating, compositeness 
possibilities.

\vspace{0.5cm}

\def\thefootnote{\arabic{footnote}}
\setcounter{footnote}{0}
\clearpage

\section{INTRODUCTION}

In previous papers we have studied the effect of Higgs boson and top
quark compositeness on $ZH$ production in gluon-gluon collision,
\cite{WLZL}, \cite{CSM}.
We have introduced the concept of Composite Standard Model (CSM)
which should preserve the structure and the main properties of the 
Standard Model (SM), at least at low energy. No anomalous coupling 
modifying the SM structures should be present, but form factors
(for example due to substructures) would affect the Higgs boson 
and top quark couplings at high energy.
We had not based our analyzes on a specific compositeness model,
for example one of those mentioned in ref.
\cite{Hcomp, comp, Hcomp2, Hcomp3, partialcomp, Hcomp4}, but on
the dependence of the observables on form factor effects.\\

We had noticed that the $gg\to ZH$ process is particularly sensitive to
the presence of form factors, because they could destroy a peculiar SM
cancellation between diagrams involving Higgs boson and top quark
couplings.
But we had also shown that this cancellation can be recovered provided that
a special relation between form factors is satisfied. We considered this
relation as a specific CSM property.\\
This $gg\to ZH$ process is therefore very interesting for revealing a violation
of the SM prediction but an amplitude analysis confirming the origin
of such an effect seems difficult to do in hadronic collisions.\\

This is why, in the present paper, we make an analysis of
$\gamma\gamma\to ZH$, the similar
process appearing in photon-photon collisions.\\
The possibility of high energy photon-photon collisions has been
considered since a long time, \cite{gammagamma1}; for a recent review
see \cite{gammagamma2}.\\
Its very interesting feature for our purpose is the possibility of working with    
polarized beams; this would allow to make an amplitude analysis
and to check what happens with the above mentioned cancellation.\\
This will be essential for choosing among the various compositeness 
possibilities, for example either with only $t_R$  
or with both $t_L$ and $t_R$ compositeness, and with or without the CSM constraint
relating it to Higgs compositeness.\\

Contents: In Section 2 we recall the definitions and the SM properties 
of the amplitudes and observables of the $\gamma\gamma\to ZH$ process
with polarized beams. In Section 3 we define the form factors, with or without
the CSM constraint, that would affect the amplitudes. 
Illustrations showing their effect on the observables are given in Section 4, 
and the conclusion and outlook in Section 5.\\

\section{AMPLITUDES AND OBSERVABLES OF THE $\gamma\gamma\to ZH$ PROCESS}

At one loop order the SM amplitudes of the $\gamma\gamma\to ZH$ process
arise from the charged fermion (leptons and quarks) loop diagrams depicted in Fig.1;
triangle diagrams (a),(b) and box diagrams (c), (with gluon-gluon symmetrization);
see  \cite{gagazh} and older references therein.\\

With CP conservation and Bose statistics the helicity amplitudes  $F_{\lambda,\lambda',\tau}$
for $\gamma,\gamma,Z$ helicities $\lambda,\lambda' =\pm{1\over2}$ and $\tau=0,\pm1$
satisfy the following relations
\bq
F_{---}=F_{+++} ~~~~F_{--0}=-F_{++0}~~~~ F_{--+}=F_{++-} 
\eq
\bq
F_{-+-}=F_{+-+}~~~~  F_{-+0}=-F_{+-0}~~~~F_{-++}=F_{+--} 
\eq

\bq
F_{\lambda,\lambda',\tau}(\theta)=(-1)^{\tau}F_{\lambda',\lambda,\tau}(\pi-\theta)
\eq
The polarized cross section has the following structure 

\bqa
{d\sigma(\gamma \gamma \to ZH)
\over d\cos\theta}&=& 
{d{\sigma}_0\over d\cos\theta}
+\langle \xi_2 \xi^\prime_2 \rangle {d{\sigma}_{22}\over d\cos\theta}
\nonumber\\
&&
-\langle\xi_3\rangle  \,{d{\sigma}_{3}\over d\cos\theta}\cos2\phi
-\langle\xi_3^ \prime\rangle {d\sigma_3^\prime\over d\cos\theta}\cos2\phi^\prime
\nonumber\\
&&
+\langle\xi_3 \xi_3^\prime\rangle
\Big [~{d{\sigma}_{33}\over d\cos\theta}
\cos (2[\phi+\phi^\prime])
+{d{\sigma}^\prime_{33}\over
d\cos\theta^*}\cos (2[\phi- \phi^\prime])\Big ]
\nonumber\\
&&+ \langle\xi_2 \xi_3^\prime\rangle
{d{\sigma}_{23}\over d\cos\theta} \sin 2 \phi
- \langle\xi_3 \xi^\prime_2\rangle
{d{\sigma^\prime}_{23}\over d\cos\theta}\sin 2\phi^\prime
 \ \ .
\label{sigpol}
\eqa
where
$(\xi_2, \xi^\prime_2)$, $(\xi_3,~ \xi^\prime_3)$
and $(\phi,~ \phi^\prime)$ describe respectively
 the average helicities, transverse
polarizations and azimuthal angles of the two photons
with their density matrices

\bq
\rho={1\over2}
\begin{pmatrix}
1+\xi_2 & -\xi_3 e^{-2i\phi}\\
-\xi_3 e^{2i\phi} & 1-\xi_2 
\end{pmatrix}~~~~
\rho'={1\over2}
\begin{pmatrix}
1+\xi'_2 & -\xi'_3 e^{2i\phi'}\\
-\xi'_3 e^{-2i\phi'} & 1-\xi'_2 
\end{pmatrix}
\eq

Using the above CP conservation relations the $\sigma_n$-quantities in (\ref{sigpol})
are defined as
\bqa
 {d \sigma_0\over d\cos\theta}&& =
\left ({\beta\over64\pi s}\right )
\sum_{\tau} [~| F_{++\tau}|^2
+| F_{+-\tau}|^2~] ~ 
\eqa
\bqa
 {d{\sigma}_{22}\over d\cos\theta}&& =
\left ({\beta\over64\pi s}\right )
\sum_{\tau} [~| F_{++\tau}|^2
-| F_{+-\tau}|^2~] 
\eqa
\bqa
{d{\sigma}_{3} \over d\cos\theta} && =
\left ({\beta\over32\pi s}\right )  \sum_{\tau}
Re [F_{++\tau}F^*_{-+\tau} ]
\eqa
\bqa
 {d{\sigma^\prime}_{3}\over d\cos\theta} && =
\left ({\beta\over32\pi s}\right ) \sum_{\tau}
Re [F_{++\tau}F^*_{+-\tau} ]
\eqa
\bqa
{d \sigma_{33} \over d\cos\theta}& & =
\left ({\beta\over 64\pi s}\right )\sum_{\tau}
Re[ F_{+-\tau} F^*_{-+\tau}~]
\eqa
\bqa
{d{\sigma^\prime}_{33}\over d\cos\theta} &&=
\left ({\beta\over 64\pi s}\right )\sum_{\tau}
Re[F_{++\tau}F^*_{--\tau}~]
\eqa
\bqa
 {d{\sigma}_{23}\over d\cos\theta}& &=
  \left ({\beta\over 32\pi s}\right ) \sum_{\tau}
  Im [
 F_{++\tau}F^*_{+-\tau} ]
\eqa
\bqa
{d{\sigma}_{23}^\prime \over d\cos\theta}&&
=  \left ({\beta\over 32\pi s}\right ) \sum_{\tau}
 Im [F_{++\tau}F^*_{-+\tau}]
\eqa
with
\bqa
\beta&= &\sqrt{\Big [1-{(m_Z-m_H)^2\over s}\Big]
\Big[1-{(m_Z+m_H)^2\over s}\Big ]} ~~~,
\eqa

Note that ${d \sigma_0/ d\cos\theta}$, ${d \sigma_{22}/ d\cos\theta}$,
${d \sigma_{33}/ d\cos\theta}$ and
${{d \sigma^\prime}_{33}/ d\cos\theta}$ are symmetric
 under the interchange
\[
\theta   \leftrightarrow \pi-\theta ~~~,
\]
whereas
\bqa
\frac {d \sigma_3}{ d\cos\theta}\Big |_\theta & =  &
\frac {d \sigma^\prime_3}{ d\cos\theta}\Big |_{\pi-\theta} ~~ ,\nonumber \\
\frac{ d \sigma_{23}}{ d\cos\theta }\Big |_\theta   & = &
 ~ \frac{d \sigma^\prime_{23}}{d\cos\theta}\Big |_{\pi-\theta} ~~ . \nonumber
\eqa

As in the $gg\to ZH$ case the leading contributions come from top quark 
loops in (a) and (c)
whereas the (b) contribution is much smaller. This smallness arises from
the mass suppressed $ZZH$ coupling appearing in (b) as compared to the $G^0ZH$ 
one in (a) which has in addition the ($m_t/m_W$) enhancement due to the
$G^0tt$ coupling.\\ 

The triangle diagrams (a) and (b) only contribute to the $F_{\pm\pm0}$
amplitudes whereas the boxes (c) contribute to all amplitudes.\\

The real and imaginary parts of the resulting SM amplitudes are shown in Fig.2a,b.\\
One can see that the helicity concerving (HC) amplitudes $F_{\pm\mp0}$ 
finally dominate over the helicity violating (HV) ones at high energy
in agreement with the HC rule \cite{hc}.\\
The remarkable cancellation between contributions of diagrams (a) 
and (c) is illustrated in Fig.3a,b.\\

The energy and angular dependences of the various $\sigma_n$ elements 
of the polarized cross section are shown in Fig.4-5.
As in the $gg\to ZH$ case the rate of longitudinal Z production is
always very large, apart from local energy and angular fluctuations. 
We only illustrate its shapes for $\sigma_0$ when looking at the
form factor effects in Fig.8a,b.\\ 

All these behaviours are globally similar to those of the $gg\to ZH$ process.
Only small differences appear at low energy where the effect of light 
lepton and quark loops is not totally negligible.\\

\section{CSM CONSTRAINTS}

Let us first recall the procedure used in \cite{CSM} for introducing
this CSM concept in the previous analysis of the $gg\to ZH$ process.\\

We affect form factors to each coupling appearing in the one loop
diagrams of this process (see Fig.1) and involving an Higgs boson
and/or a top quark.  This means five arbitrary form factors chosen as
$F_{G^0Z_LH}(s)=F_{ZZ_LH}(s),F_{Htt}(s),F_{Gtt}(s),F_{tR}(s),F_{tL}(s)$.\\

Incidently we note that photon-$tt$ form factors (similarly to
gluon-$tt$ form factors) may also appear, 
but as they would occur exactly in the same way
in the triangles and in the boxes, they would not affect the structures
of the amplitudes (in particular the special cancellation mentioned in the
previous section); they would only modify the total result
by a pure normalization factor. So we do not discuss them more.\\
 
With the five arbitrary form factors listed above large effects 
are in general observed due to the destruction of the cancellation
appearing for the $F_{++0}$ amplitude between its contribution from
triangle (a) and the one from boxes (c) (see Fig.8 of \cite{CSM}).
But this cancellation can be recovered provided that the following 
\underline{CSM constraint}
is satisfied:
\bq
F_{G^0Z_LH}(s)F_{Gtt}(s)(g^Z_{tR}-g^Z_{tL})=
F_{Htt}(s)(g^Z_{tR}F_{tR}(s)-g^Z_{tL}F_{tL}(s))~~\label{CSMconsFF}
\eq

We now look at the $\gamma\gamma\to ZH$ process and the
effect of such form factors, with and without
the CSM constraint, on the observables defined in the previous section.\\

For the illustrations, as simple examples, we will use simple "test" expressions 
of the type
\bq
F(s)={(m_Z+m_H)^2+M^2\over s+M^2}
\eq
\noindent
with the new physics scale $M$ taken for example with the value of 5 TeV.\\
We use this expression for different form factors and we
treat separately the following cases:\\

The two CSM conserving cases will be\\ 

--- (a) denoted CSMtLR, refering to both $t_{L}$ and $t_{R}$ compositeness, with
$F_{tR}(s)=F_{tL}(s)\equiv F(s)$,
and $F_{G^0Z_LH}(s)=F_{Gtt}(s)=F_{Htt}(s)\equiv F(s)$ satisfying (\ref{CSMconsFF}),\\
 
--- (b) denoted CSMtR, for pure $t_R$ compositeness, $F_{tR}(s)\equiv F(s)$ and $F_{tL}(s)=1$,
and, in order to satisfy (\ref{CSMconsFF}), either\\
 $F_{Gtt}(s)=F_{Htt}(s)\equiv F(s)$
and $F_{G^0Z_LH}(s)(g^Z_{tR}-g^Z_{tL})=g^Z_{tR}F_{tR}(s)-g^Z_{tL}$\\
or\\
$F_{G^0Z_LH}(s)=F_{Htt}(s)\equiv F(s)$
and $F_{Gtt}(s)(g^Z_{tR}-g^Z_{tL})=g^Z_{tR}F_{tR}(s)-g^Z_{tL}$,\\

and the 2 CSM violating cases:\\

--- (c) denoted CSMvt, where $F_{tR}(s)$ with $M=10$ TeV, $F_{tL}(s)$ with $M=15$ TeV
and 
$F_{G^0Z_LH}(s)=F_{Htt}(s)=F_{Gtt}(s)\equiv F(s)$ with $M=5$ TeV,\\

--- (d) denoted CSMvH, no top form factor and only one  $F(s)$ form factor affecting the 
$G^0Z_LH$ and $ZZ_LH$ vertices.\\

In the next Section and Figures 6-8, we illustrate the consequences of such choices 
for the observables.\\

\section{ILLUSTRATIONS}

Fig.6a-f and  7a-f respectively illustrate the effects of the choices (a-d) of
form factors on the energy and angular dependences 
of the $\sigma_n$ cross sections corresponding to the unpolarized or to specific
polarization cases.
Fig.8a,b illustrate similarly the behaviour of the $Z_L$ rate in the
$\sigma_0$ cross section.\\

As in the gluon-gluon case we can first check
that the unpolarized $\sigma_0$ cross section is very
sensitive to the CSM violation cases CSMvt and CSMvH. The corresponding $Z_L$ rates
tend also to be closer to 1 at high energy.\\
Similar effects can be seen in the longitudinal polarization term $\sigma_{22}$.\\
The transverse polarization terms $\sigma_{3}$ (related to $\sigma'_{3}$) 
and $\sigma'_{33}$ are also very sensitive with spectacular changes of signs.\\
On the opposite the double transverse $\sigma_{33}$ term, only controlled by HC amplitudes, is not
affected by the CSMvH case, but gets some effects in the other cases.\\
The mixed longitudinal and transverse term $\sigma_{23}$ (related to $\sigma'_{23}$) involving
both HV and HC amplitudes get also notably modified by the CSM violating terms.\\

Globally the CSM conserving terms lead to weaker effects in these various cross section elements.\\

Quantitatively the above effects obviously depend on the precise expression used for the 
"test" form factor and on the value of $M$; our simple choice has no physical meaning 
but was only chosen in order to show the sensitivity of each observable.\\

Note also that the CSM violating cases producing large differences with the pure
SM prediction due to the destruction of the special cancellation in $F_{\pm\pm0}$
amplitudes could nevertheless satisfy unitarity owing to the decrease of the
form factors above the new physics scale (chosen at 5 TeV in the illustrations).\\

\section{CONCLUSIONS}

We have shown that the $\gamma\gamma\to ZH$ process, especially with polarized
photon beams, should be particularly interesting for analyzing a possible
departure from SM predictions with the aim of testing the CSM concept.\\
Indeed the observables of the $\gamma\gamma\to ZH$ process with
photon polarizations and the associated $\theta,\phi,\phi'$ dependences 
should allow to make an amplitude analysis and to identify the form factor 
structure, CSM conserving or CSM violating.\\ 
As in the $gg\to ZH$ case, the biggest effect should be found in the $F_{\pm\pm0}$
amplitude whose internal cancellation property is preserved or violated by
the corresponding form factors.\\
We have illustrated with a simple "test" shape how the various observables
react to the presence of a form factor in the involved couplings.\\
We can conclude, in the case a departure from SM prediction would be
observed in $gg\to ZH$, that the $\gamma\gamma\to ZH$ process would be
very fruitful for revealing the nature of a possible compositeness
interpretation.\\

Along this line of thought further works could be carried
with phenomenological analyzes of other processes sensitive
to such compositeness effects, with or without CSM constraint;
see for example $WW,ZZ\to t\bar t$ and ref.\cite{wwtt}.
Independently, possible theoretical studies could be done
about the meaning of the CSM concept.\\

\newpage

\begin{figure}[p]
\[
\epsfig{file=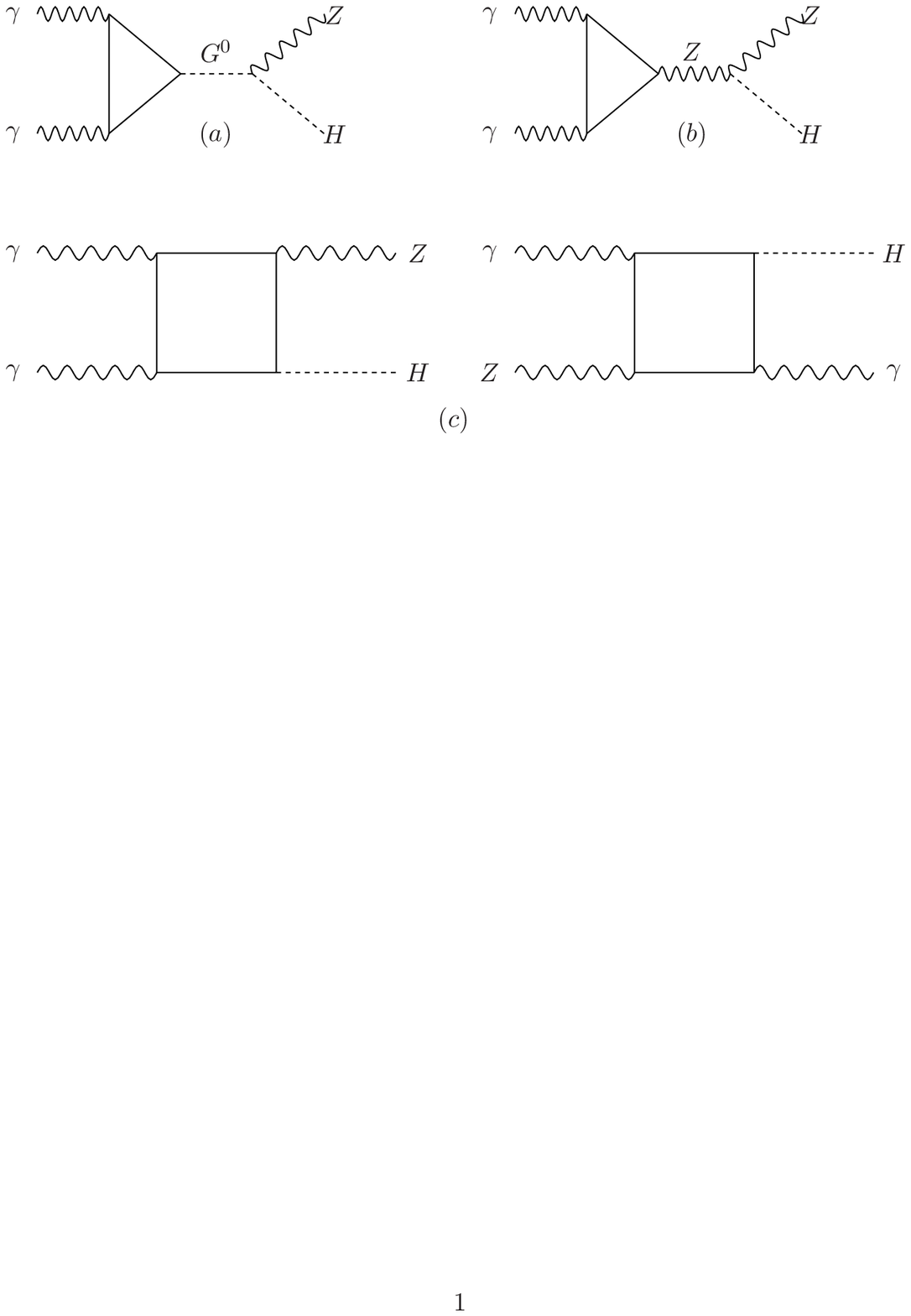, height=24.cm}
\]\\
\vspace{-15cm}
\caption[1] {The triangle and box one loop SM diagrams contributing
to the $\gamma\gamma\to ZH$ process.}
\end{figure}

\clearpage

\begin{figure}[p]
\[
\epsfig{file=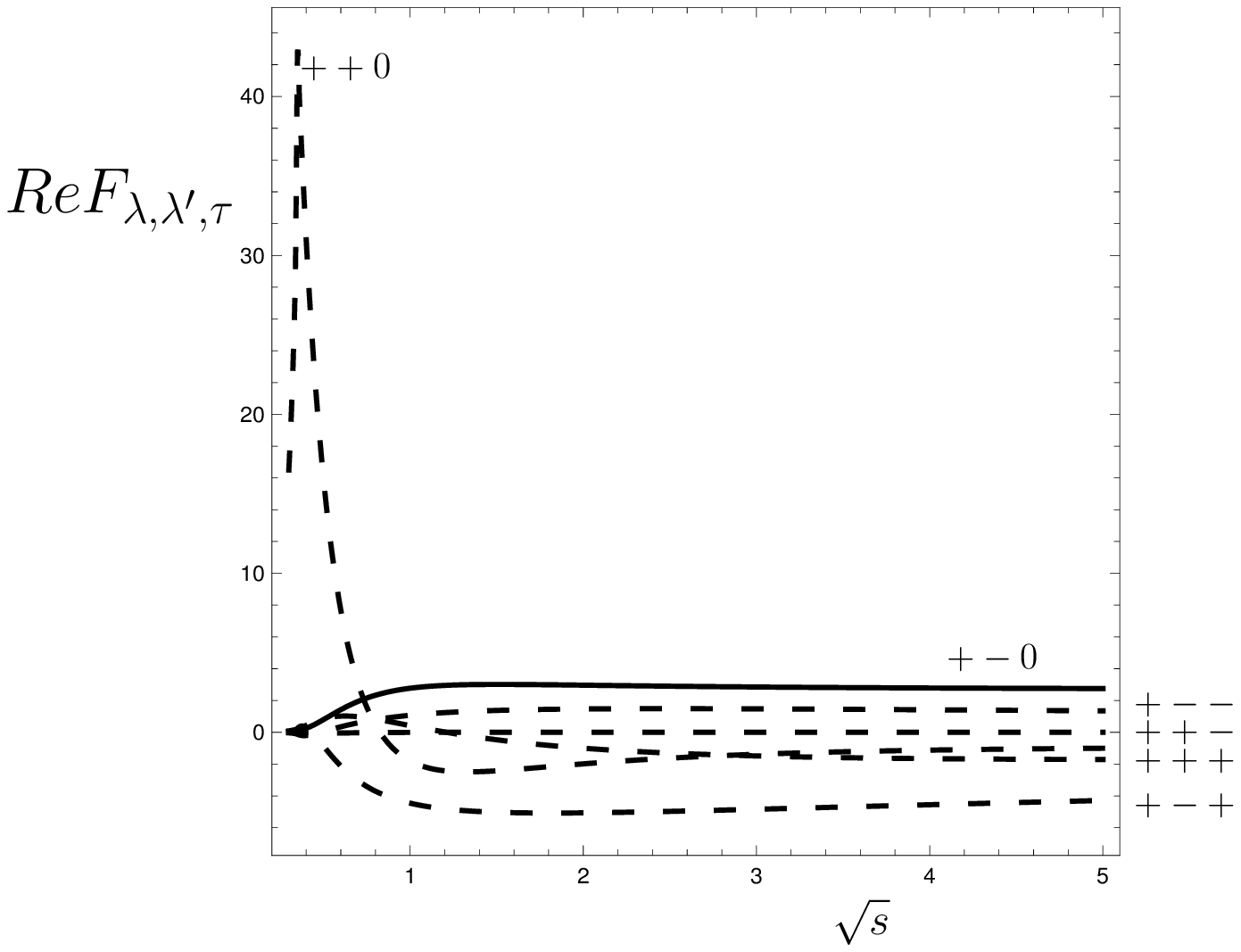, height=8.cm}
\]\\
\vspace{-1cm}
\[
\epsfig{file=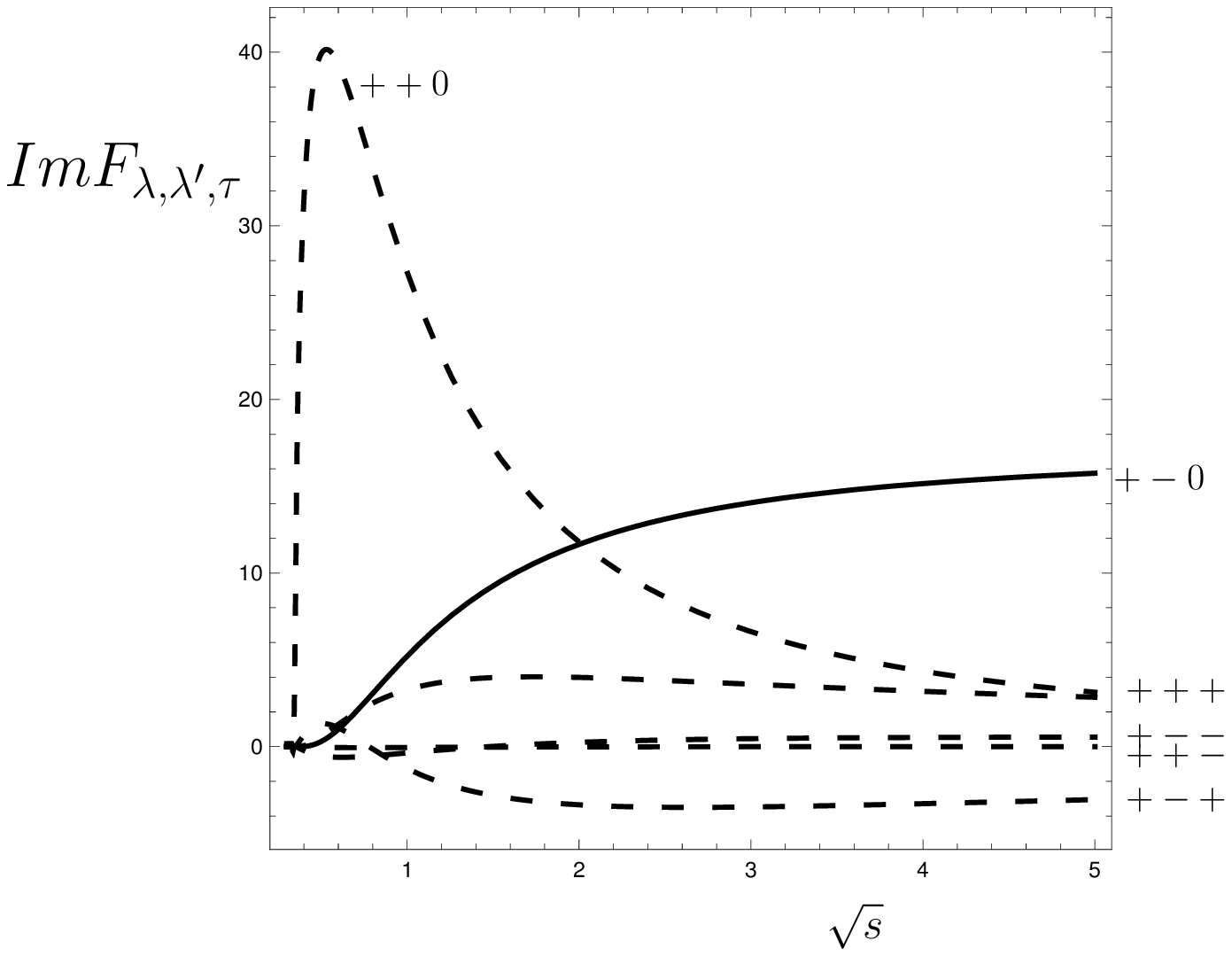, height=8.cm}
\]\\
\vspace{-1cm}
\caption[1] {Energy dependence of the real and imaginary parts of the 6 independent SM amplitudes.}
\end{figure}

\clearpage

\begin{figure}[p]
\[
\epsfig{file=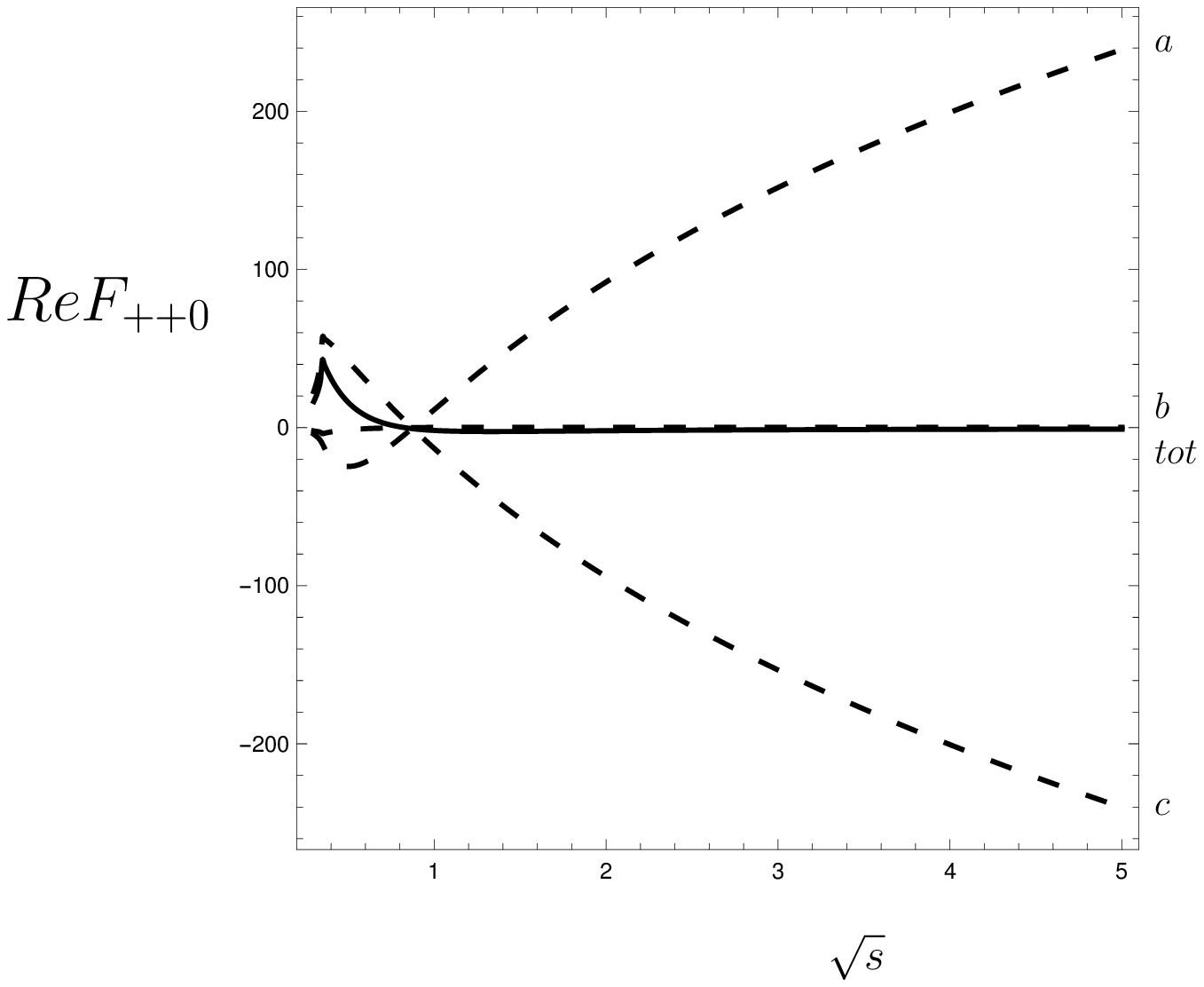, height=8.cm}
\]\\
\vspace{-1cm}
\[
\epsfig{file=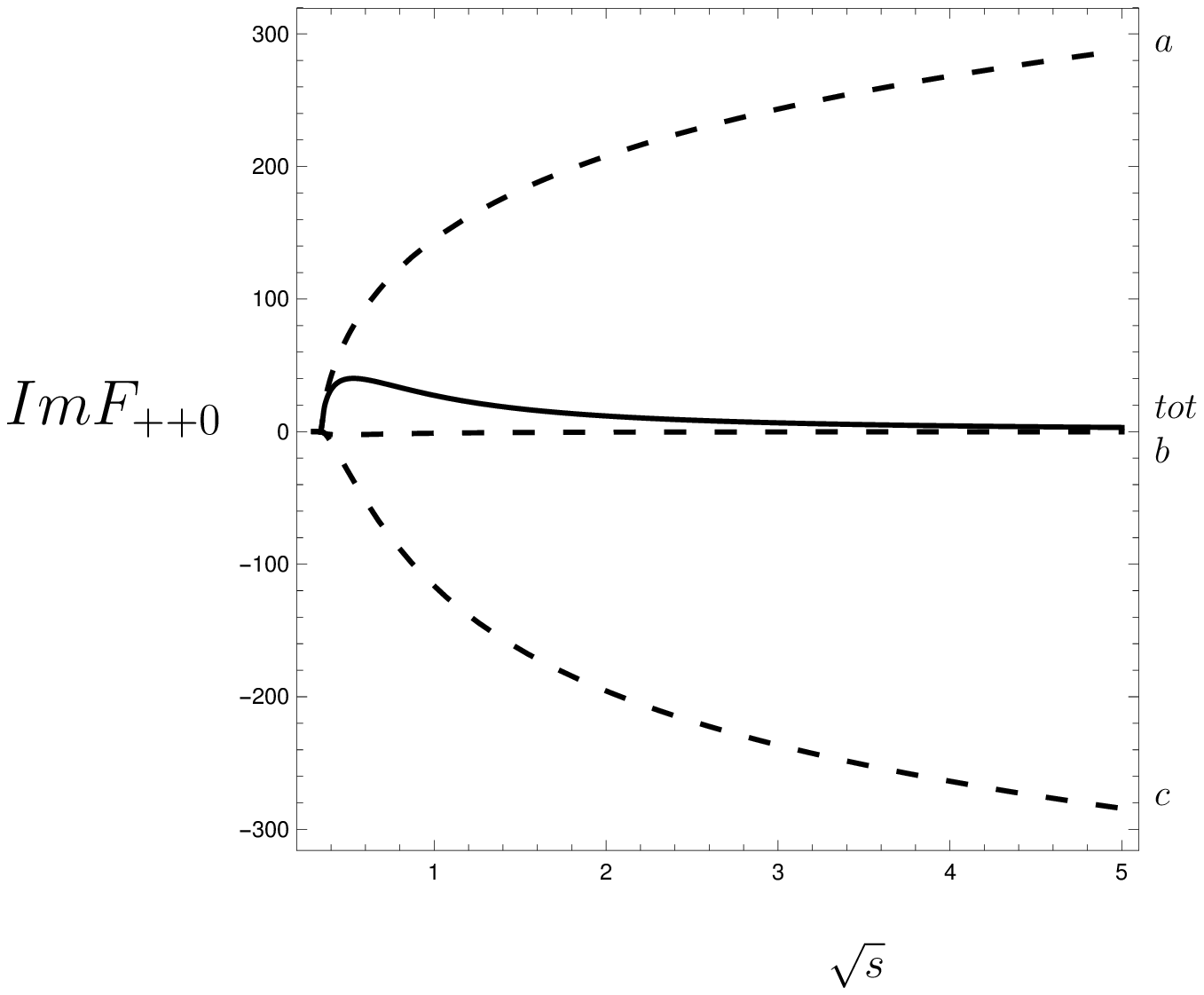, height=8.cm}
\]\\
\vspace{-1cm}
\caption[1] {Energy dependences of the real and imaginary parts of the SM contributions to the $F_{++0}$  amplitude; (a),(b),(c) diagrams and total.}
\end{figure}

\clearpage

\begin{figure}[p]
\[
\epsfig{file=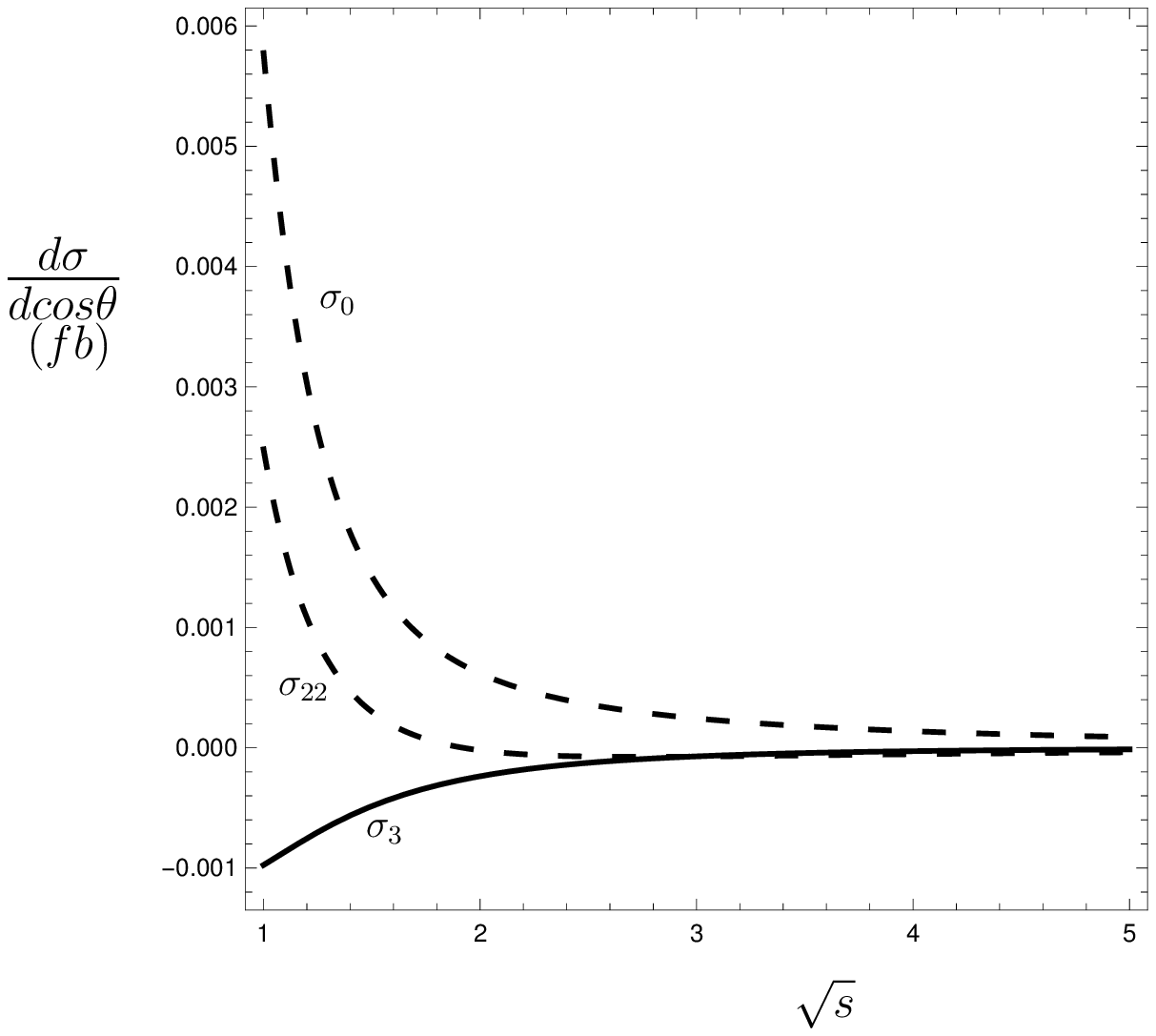, height=8.cm}
\]\\
\vspace{-1cm}
\[
\epsfig{file=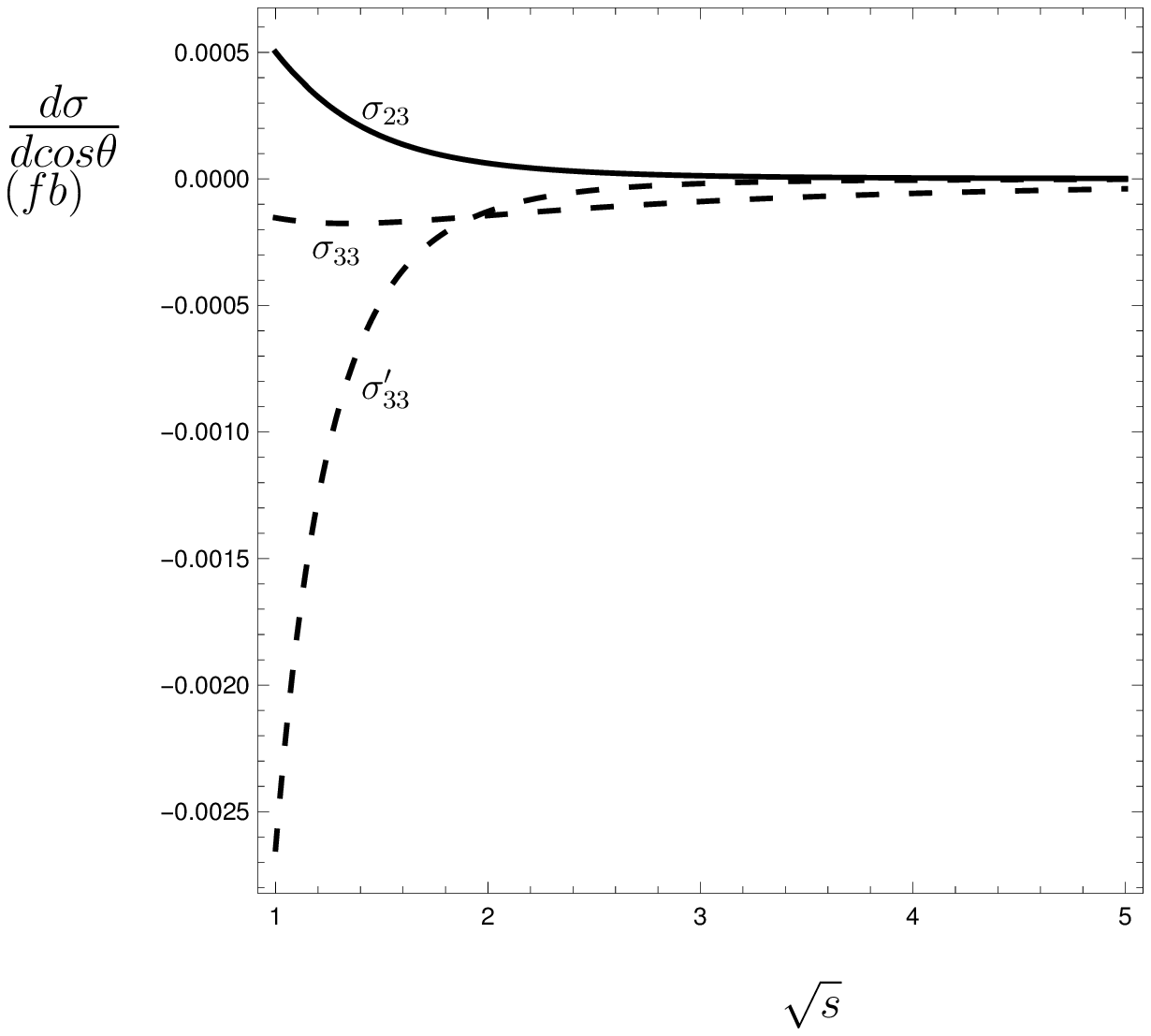, height=8.cm}
\]\\
\vspace{-1cm}
\caption[1] {Energy dependence of SM cross sections.}
\end{figure}

\clearpage
\begin{figure}[p]
\[
\epsfig{file=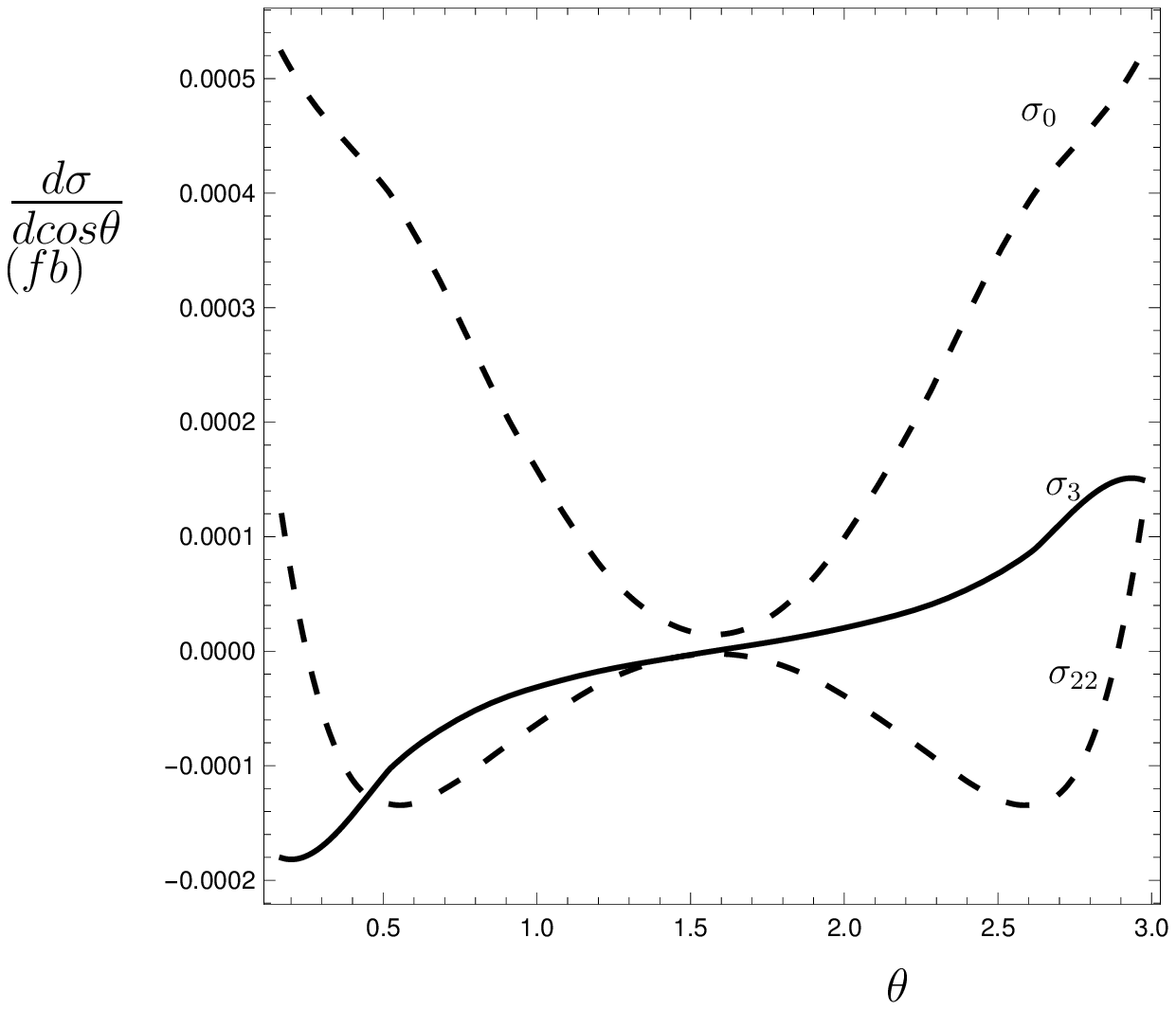, height=8.cm}
\]\\
\vspace{-1cm}
\[
\epsfig{file=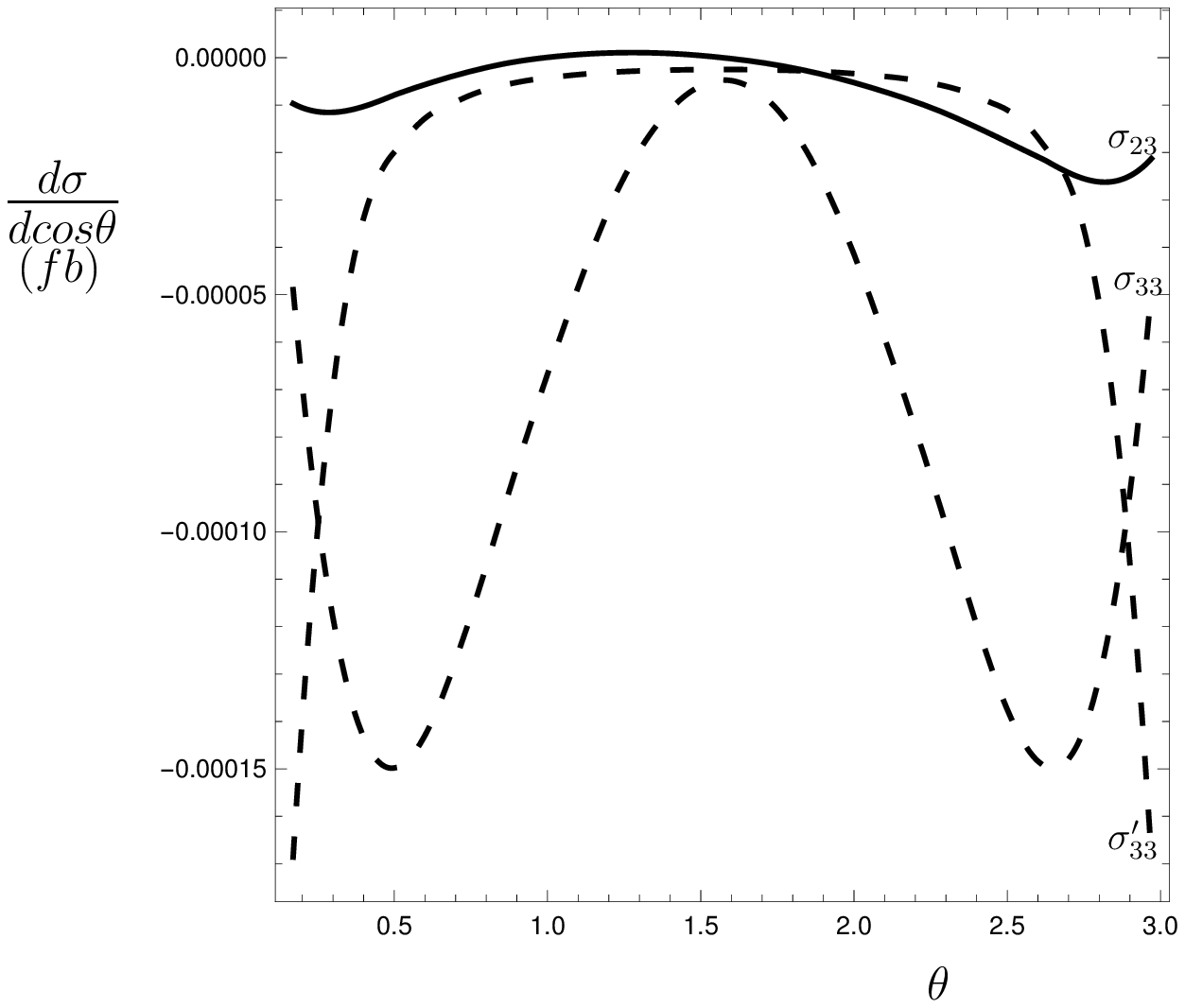, height=8.cm}
\]\\

\vspace{-1cm}
\caption[1] {Angular distributions of SM cross sections at 4 TeV.}
\end{figure}

\clearpage
\begin{figure}[p]
\[
\epsfig{file=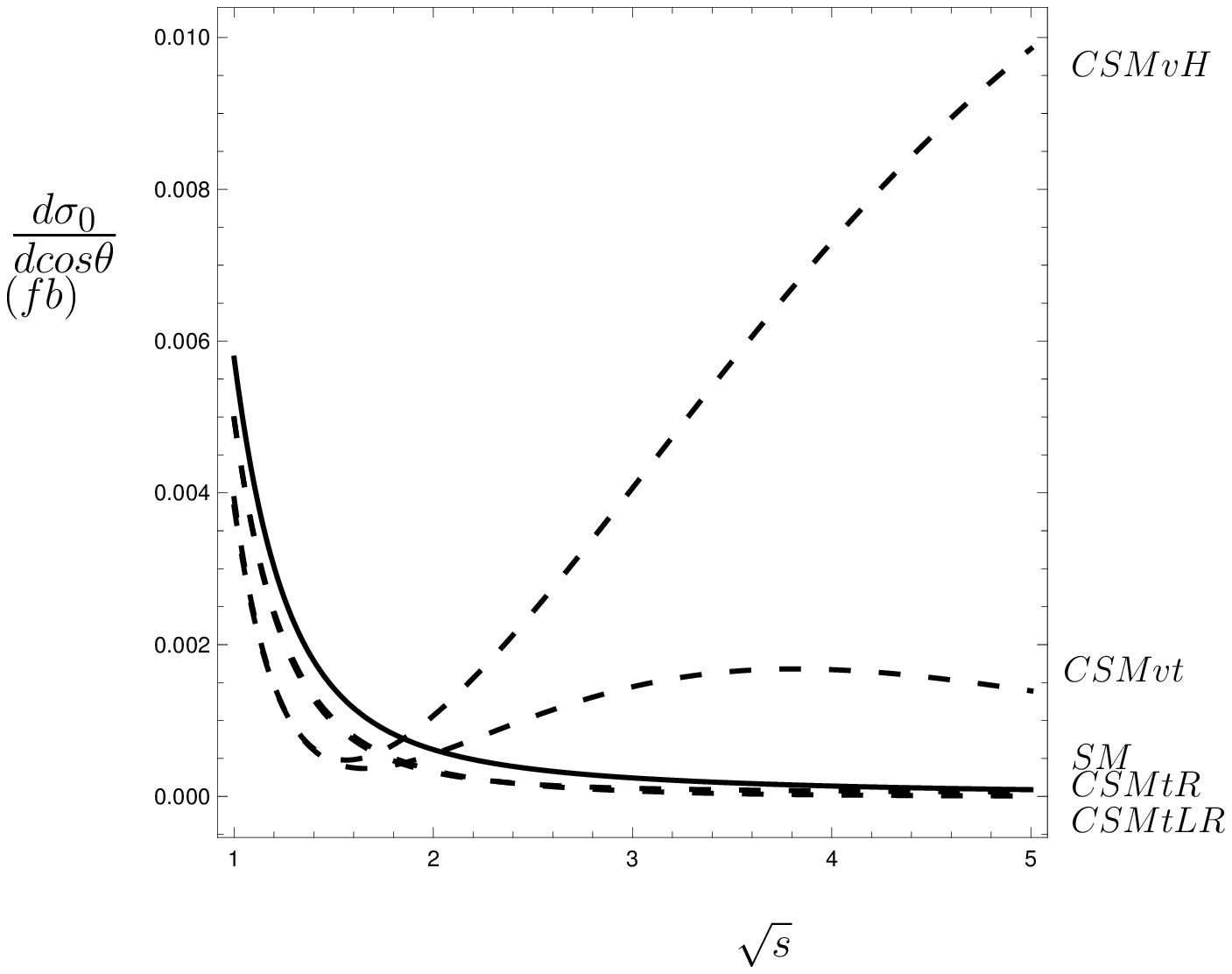, height=6.cm}
\hspace{-0.2cm}\\
\epsfig{file=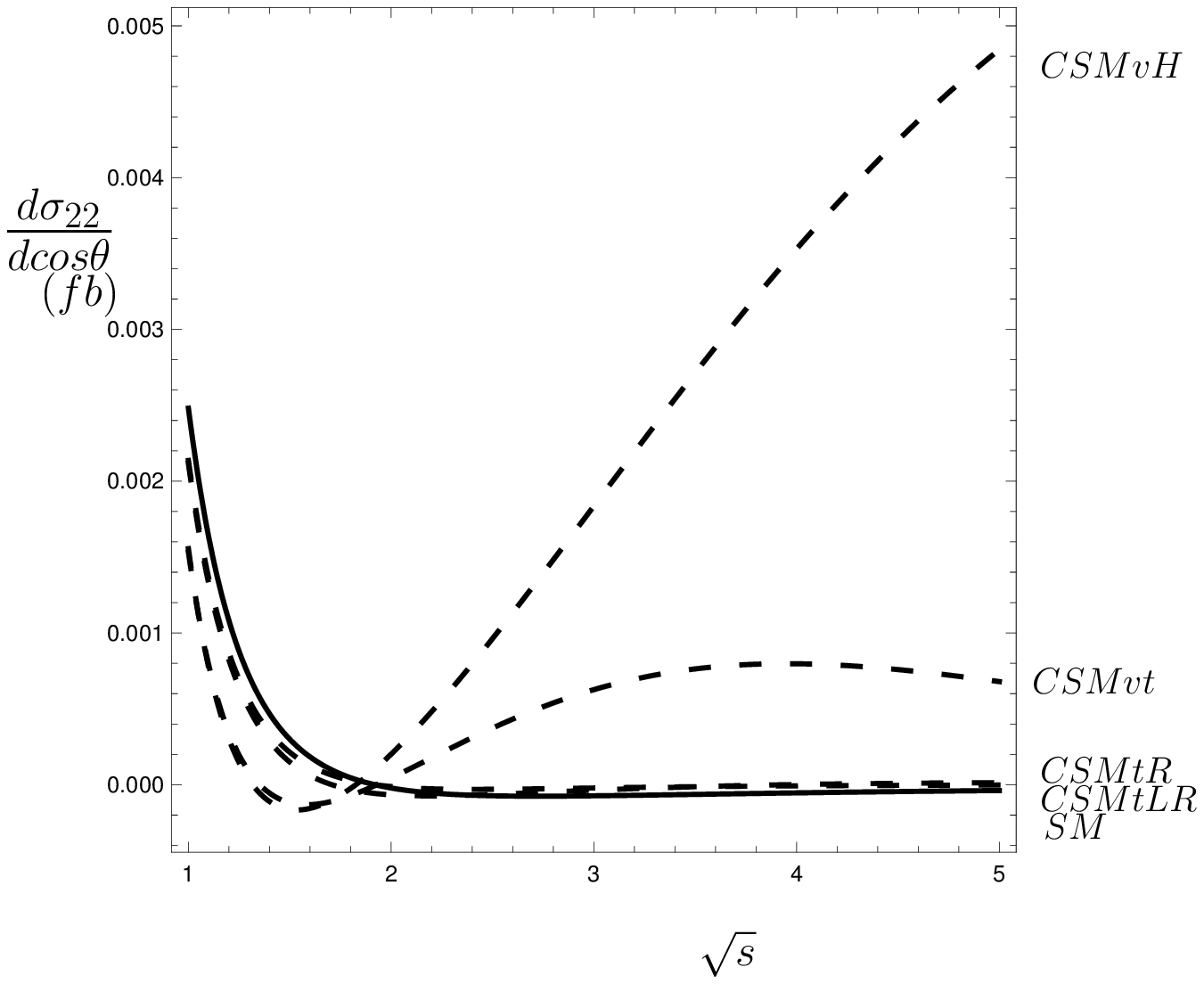, height=6.cm}
\]\\
\vspace{-1cm}
\[
\epsfig{file=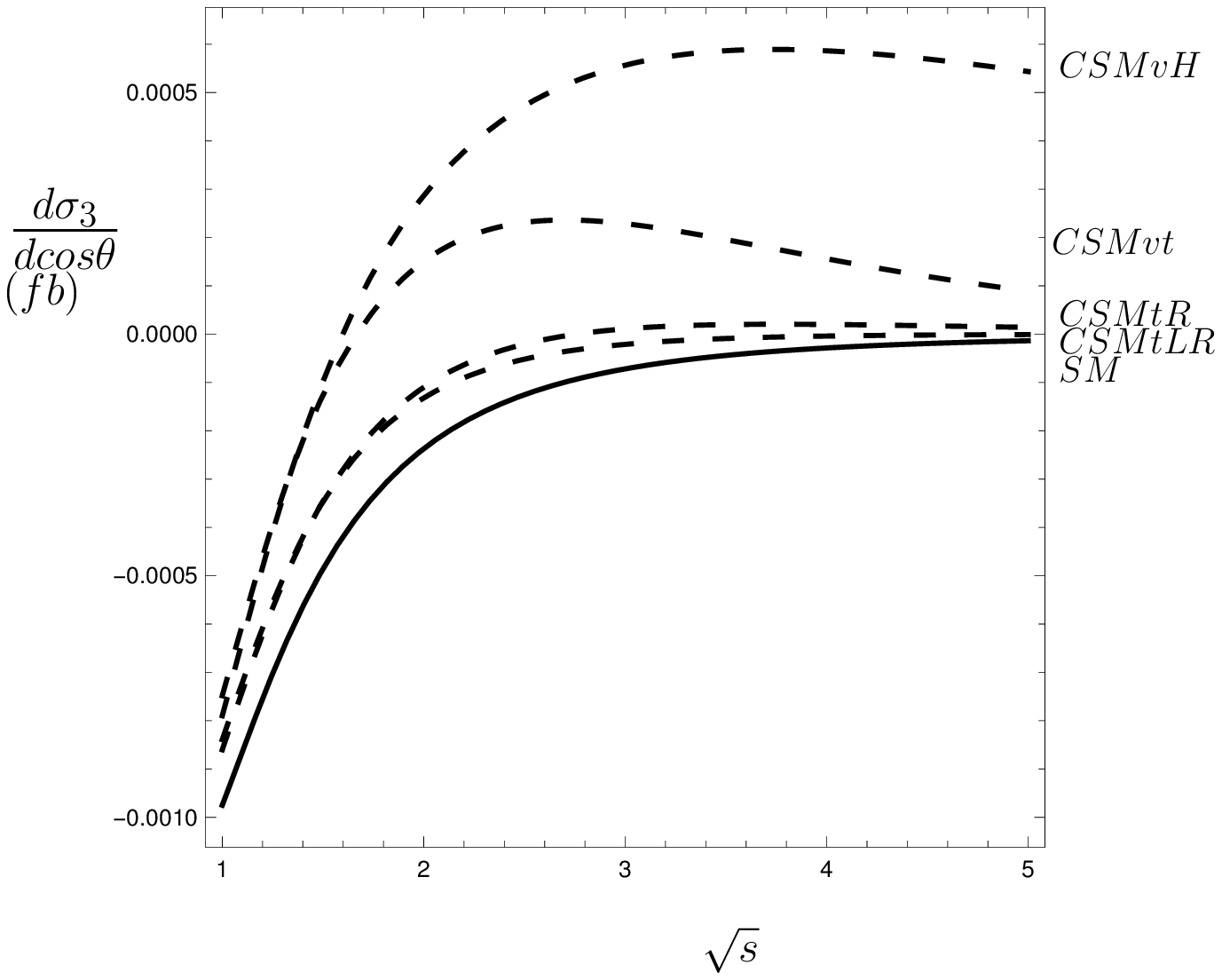, height=6.cm}
\hspace{-0.2cm}\\
\epsfig{file=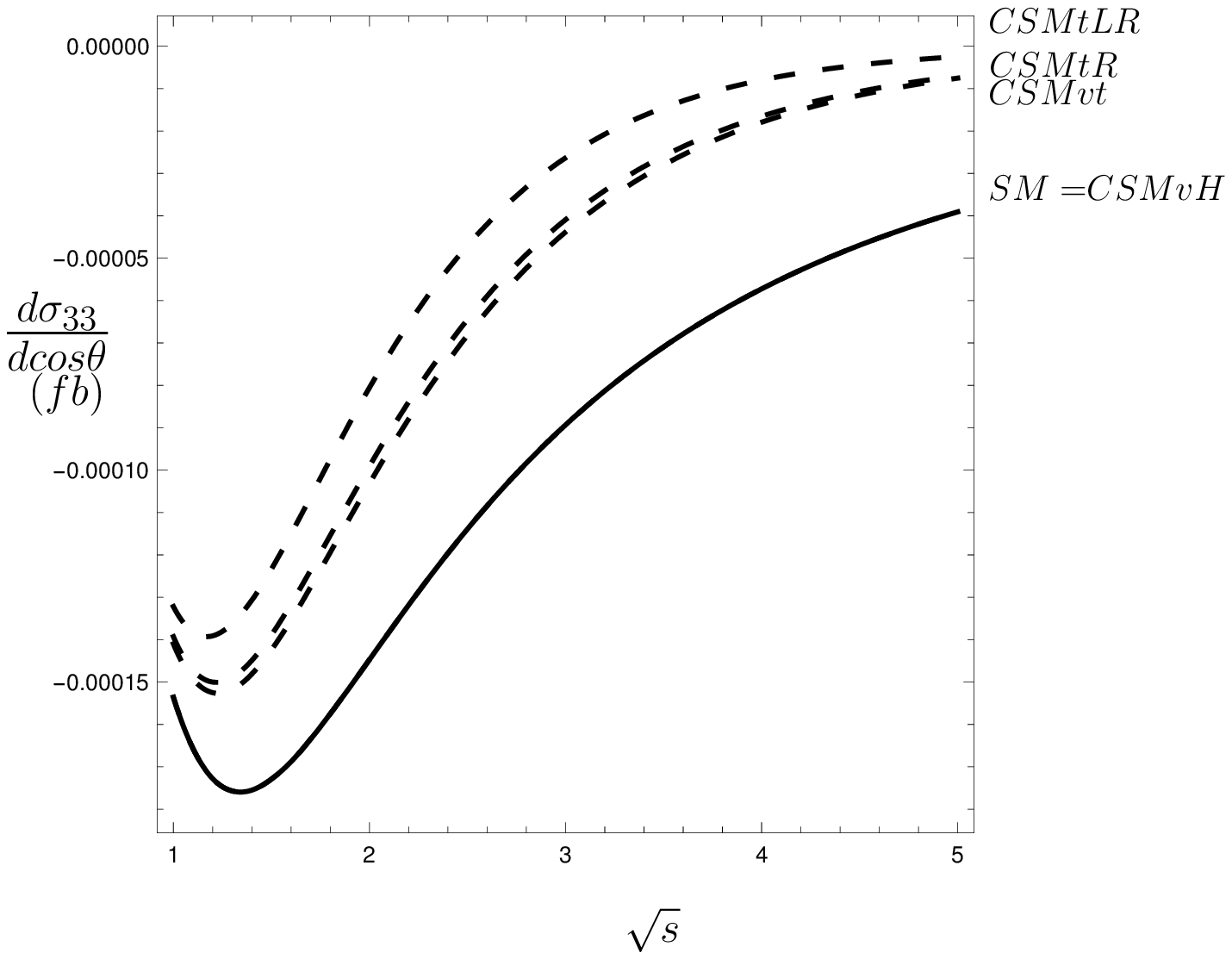, height=6.cm}
\]
\vspace{0cm}
\[
\epsfig{file=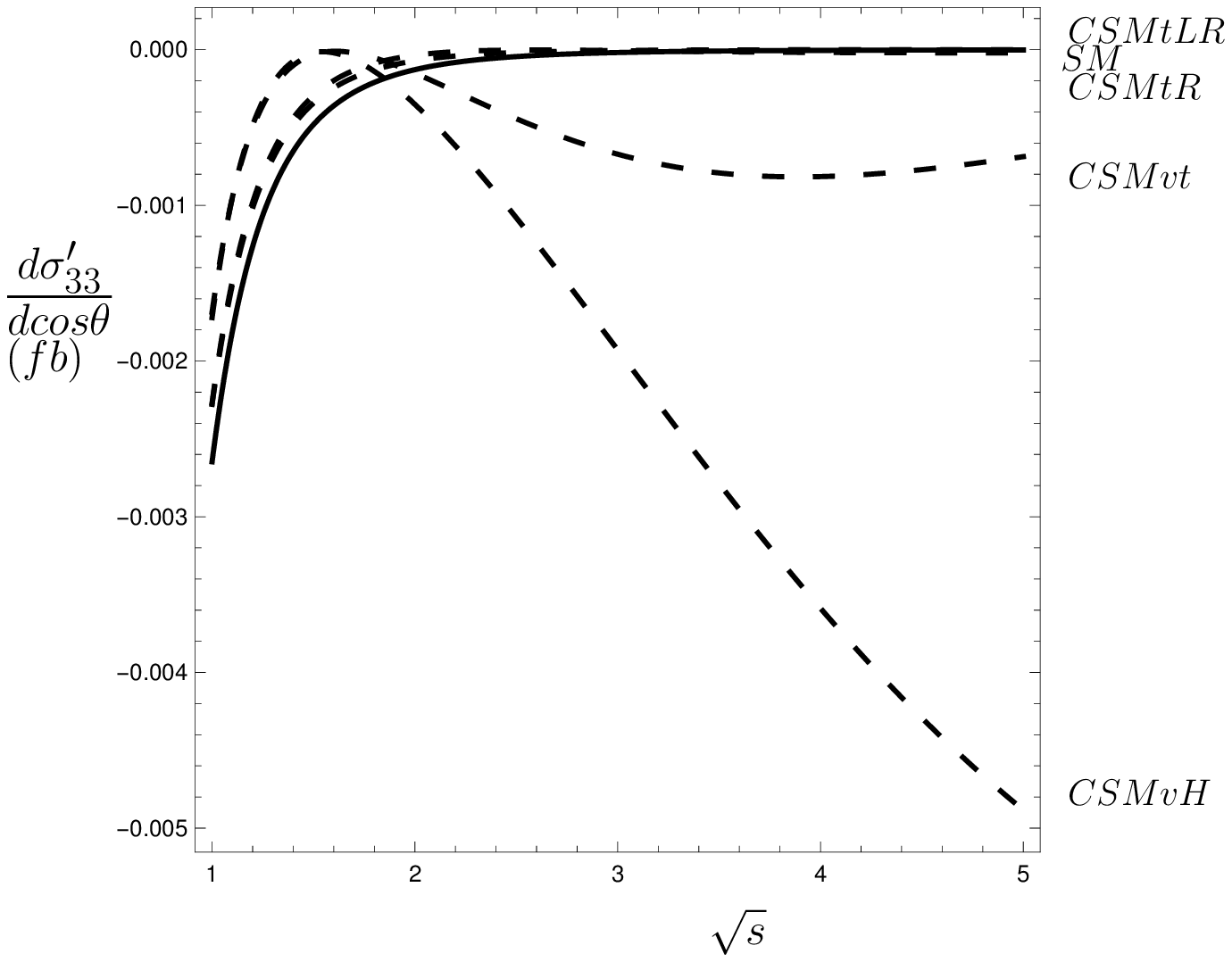, height=6.cm}
\hspace{-0.2cm}\\
\epsfig{file=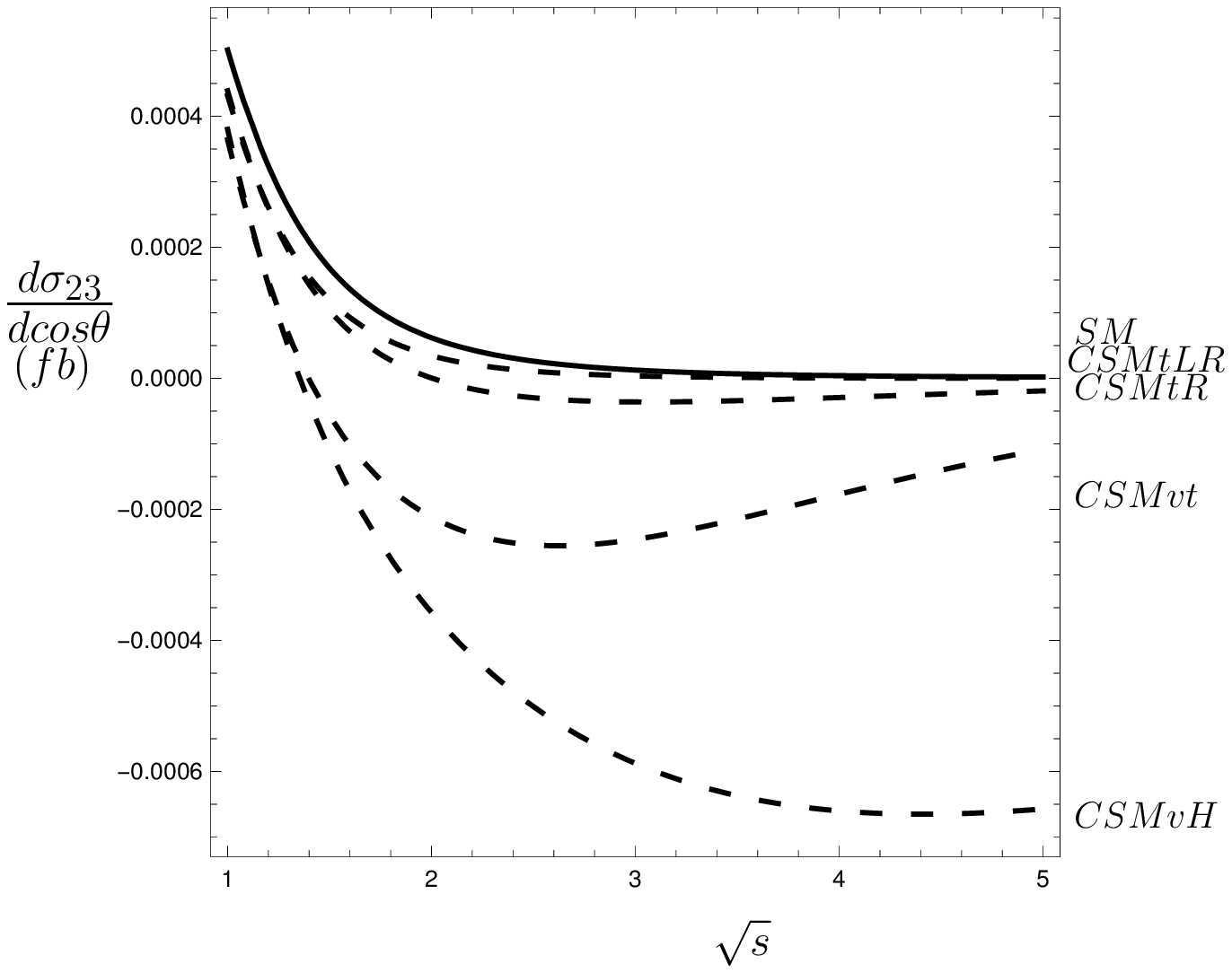, height=6.cm}
\]\\

\caption[1] {Cross sections in SM, CSMtLR, CSMtR, CSMvt, CSMvH cases.}
\end{figure}

\clearpage

\clearpage
\begin{figure}[p]
\[
\epsfig{file=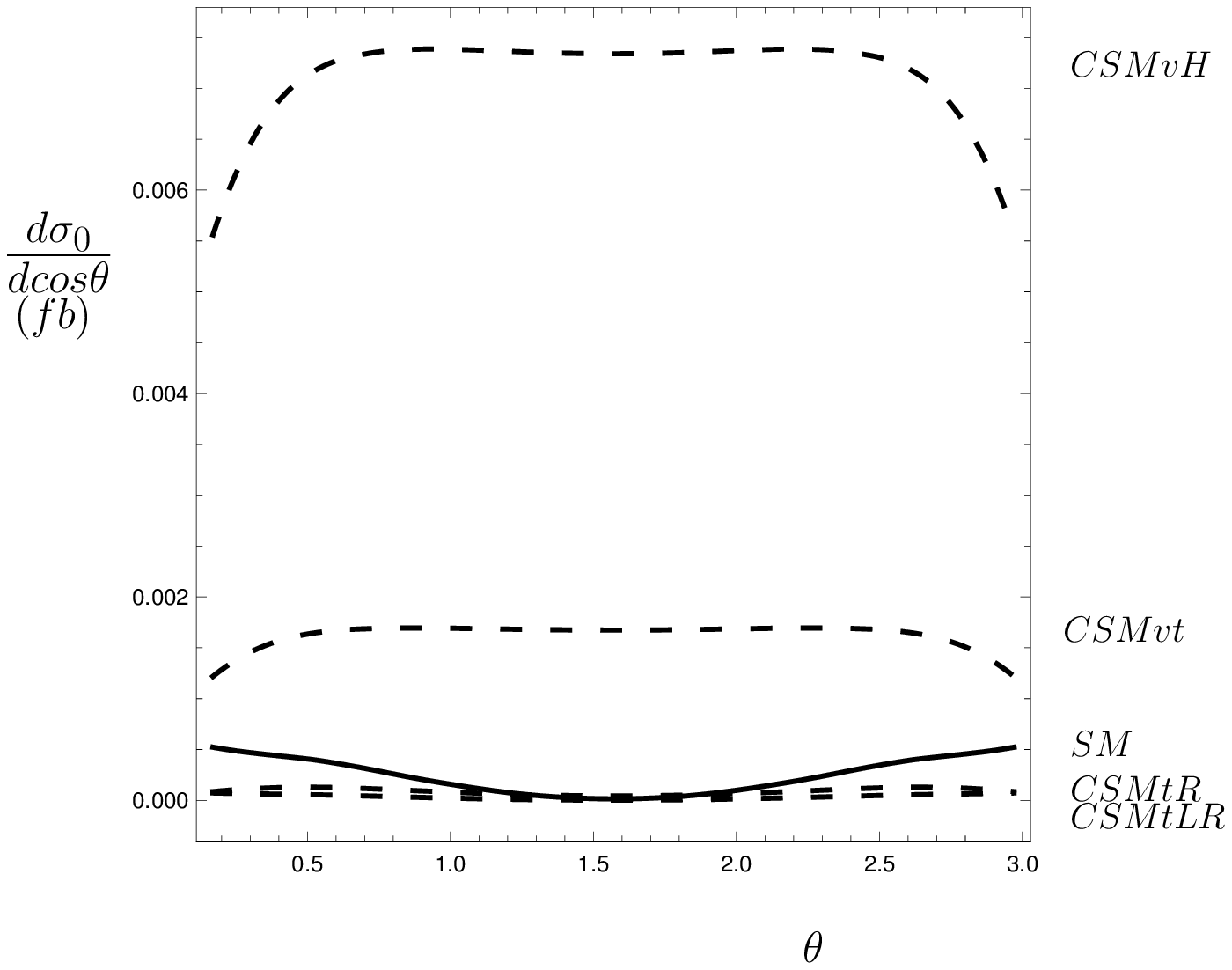, height=6.cm}
\hspace{-0.2cm}\\
\epsfig{file=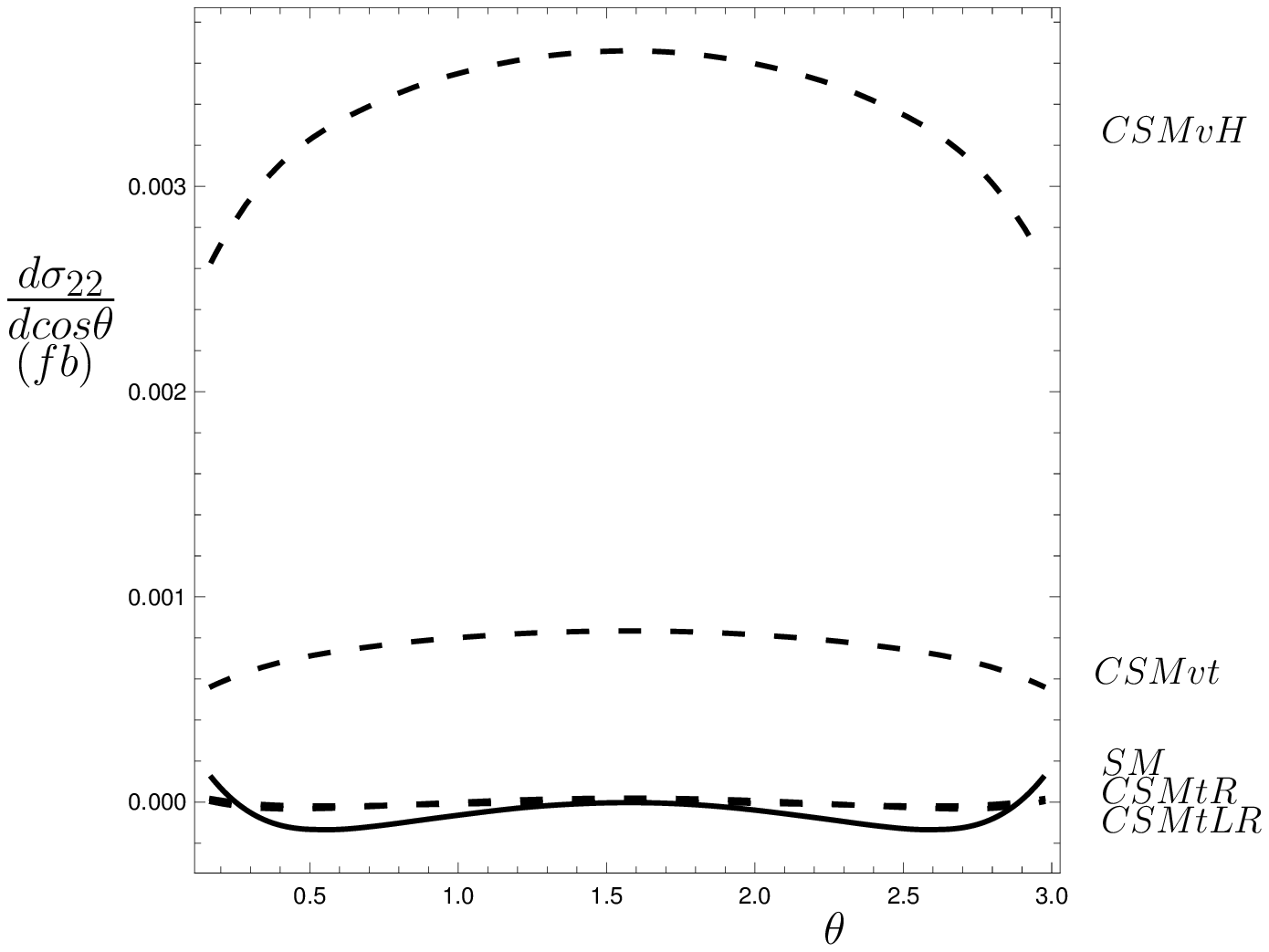, height=6.cm}
\]\\
\vspace{-1cm}
\[
\epsfig{file=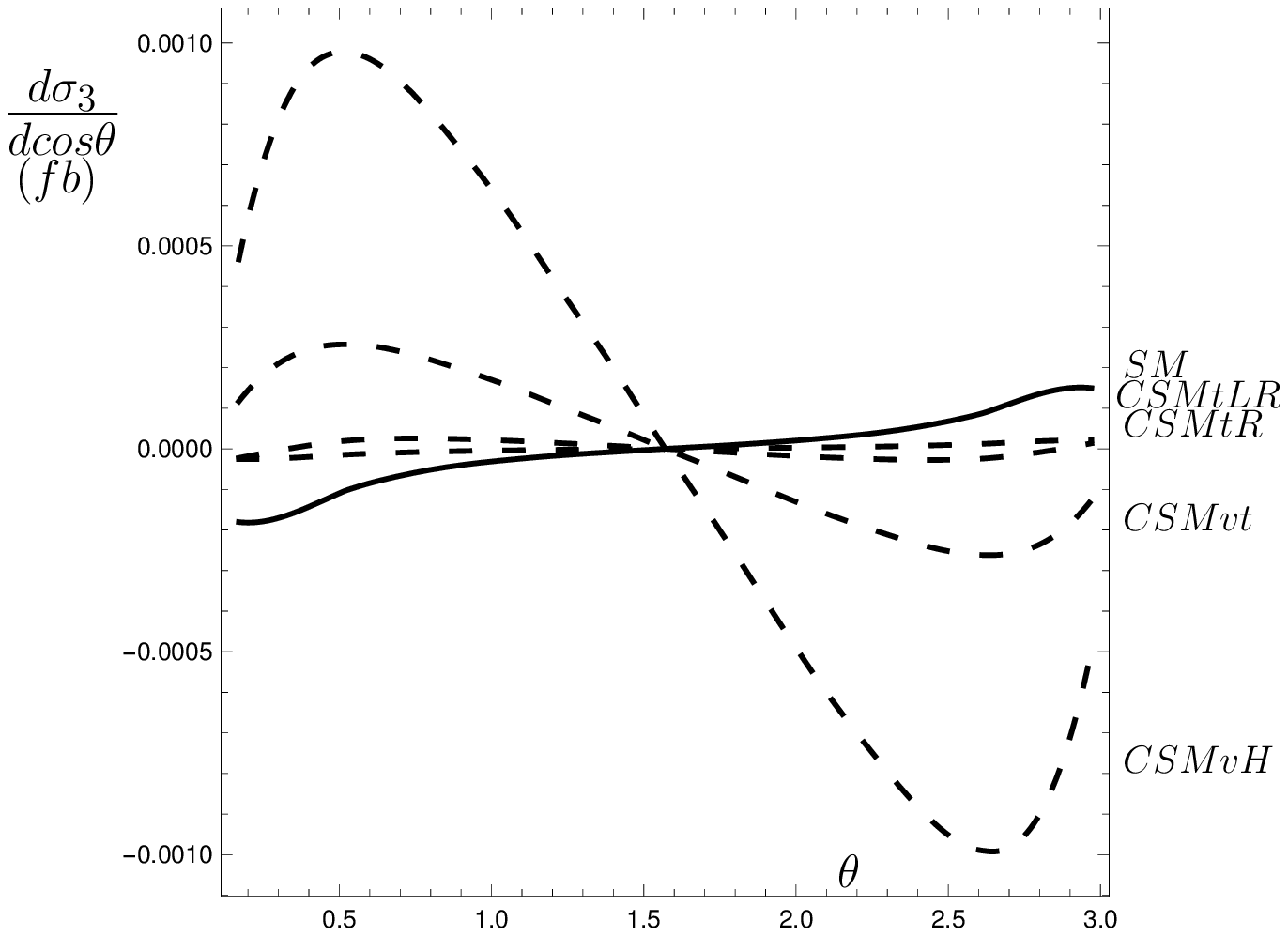, height=6.cm}
\hspace{-0.2cm}\\
\epsfig{file=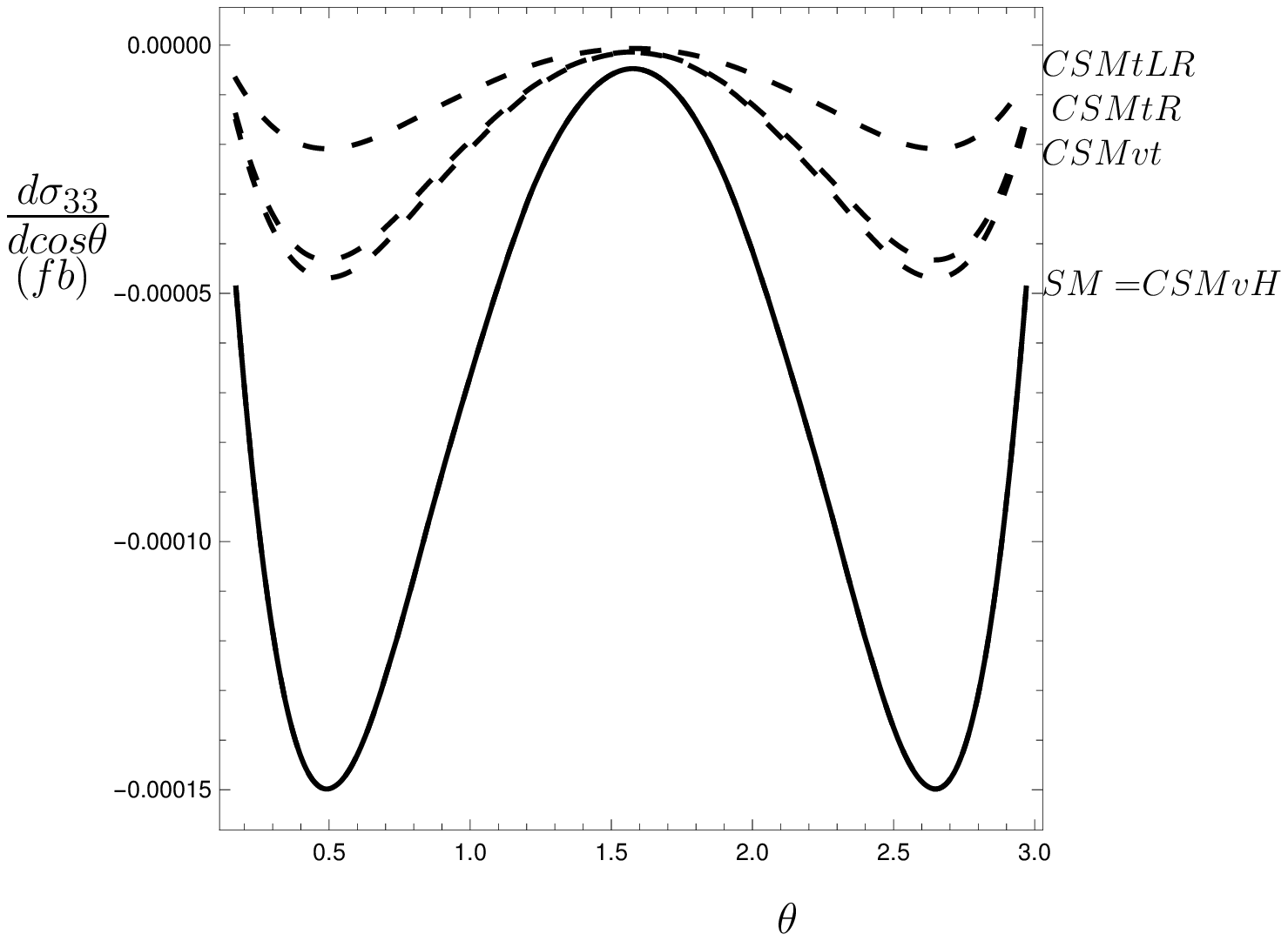, height=6.cm}
\]
\vspace{0cm}
\[
\epsfig{file=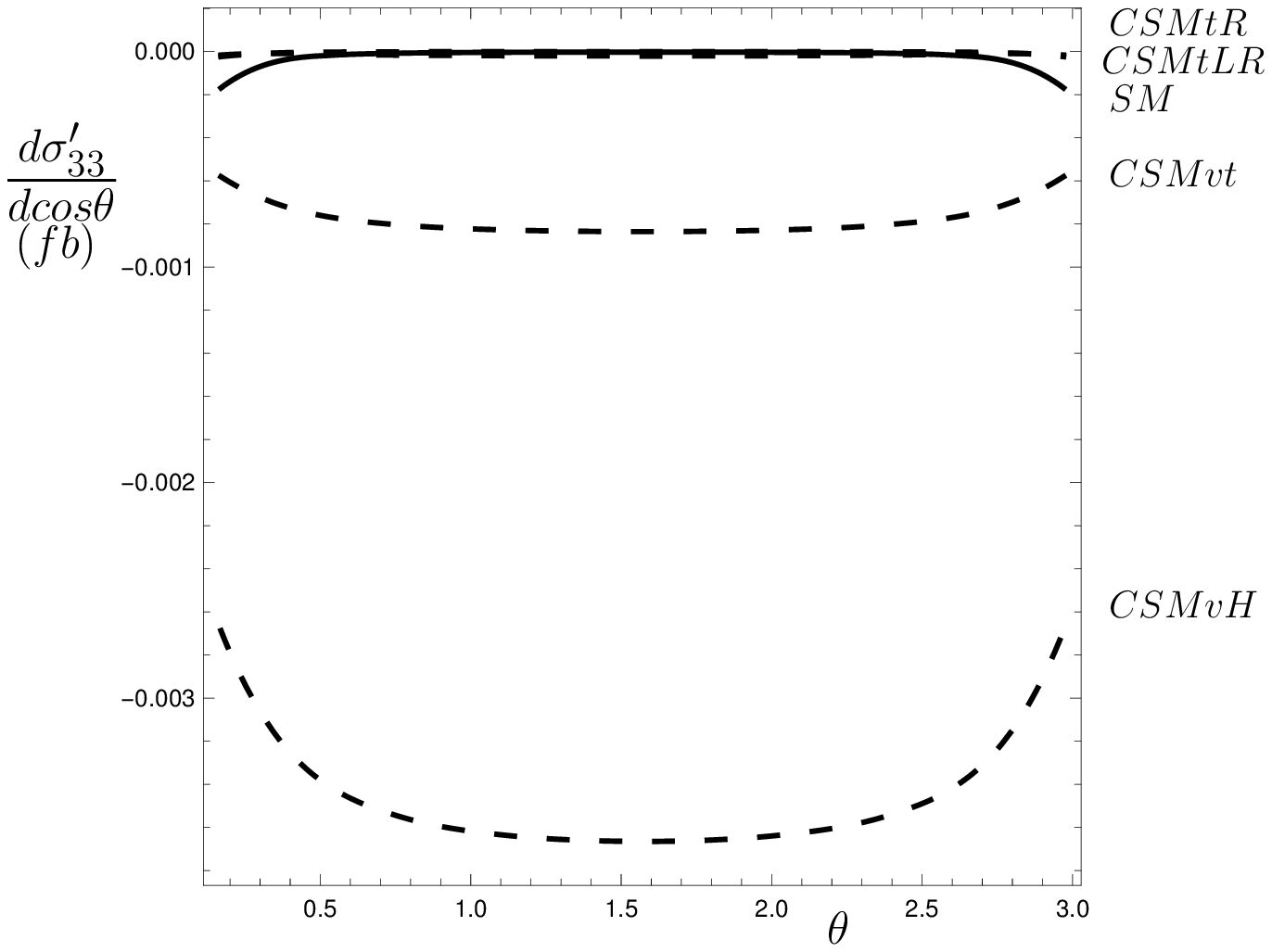, height=6.cm}
\hspace{-0.2cm}\\
\epsfig{file=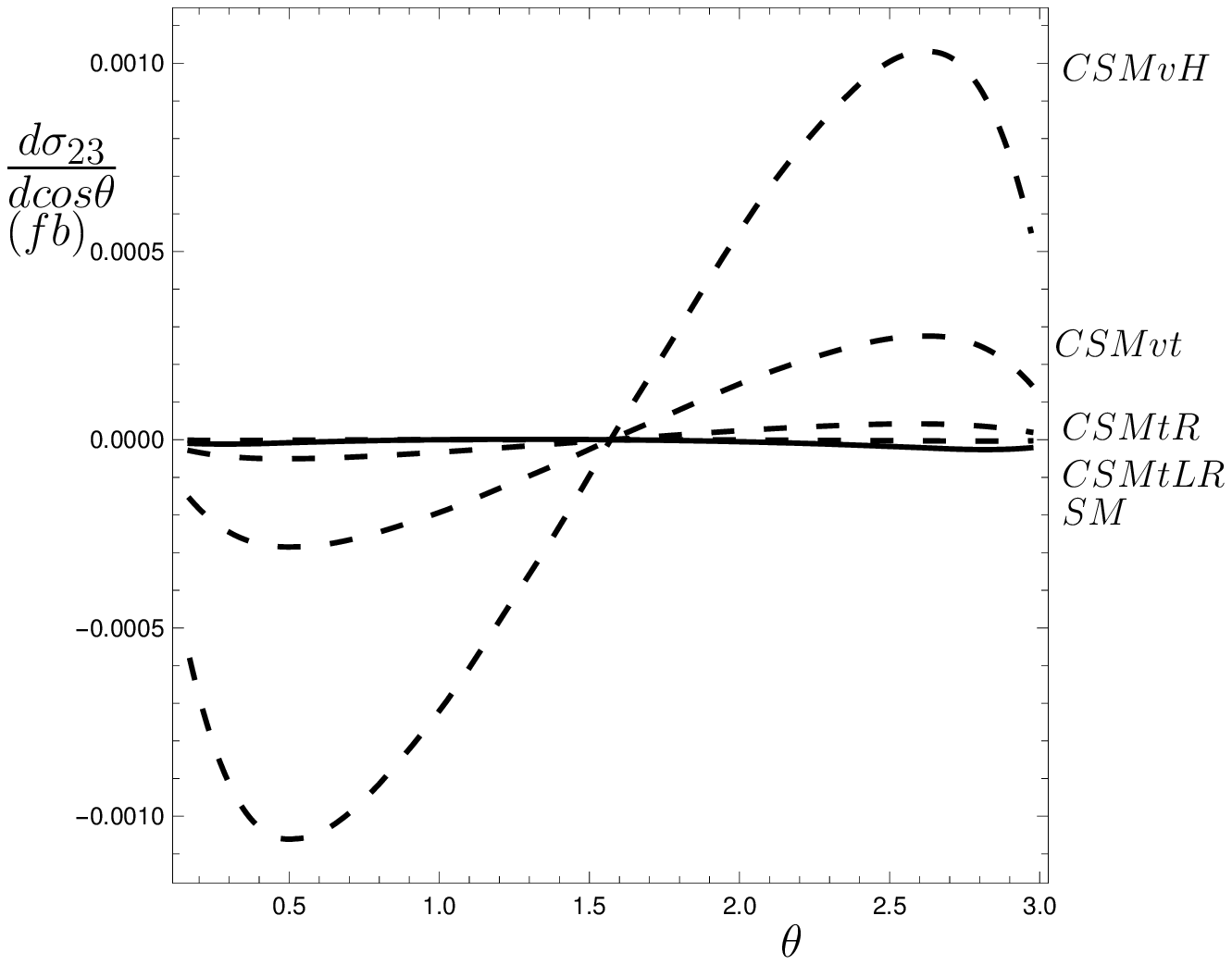, height=6.cm}
\]\\

\caption[1] {Angular distributions in SM, CSMtLR, CSMtR, CSMvt, CSMvH cases.}
\end{figure}

\clearpage

\clearpage
\begin{figure}[p]
\[
\epsfig{file=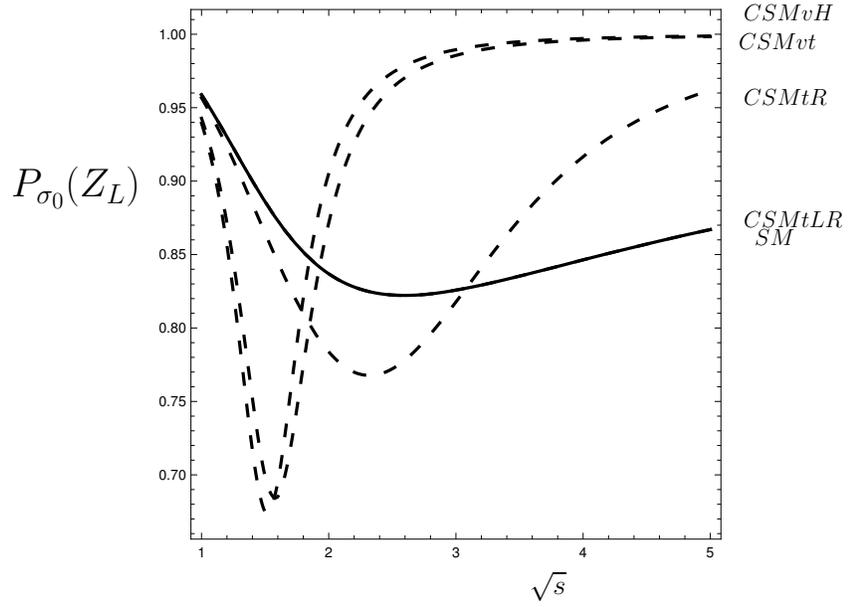, height=8.cm}
\]\\
\vspace{0cm}
\[
\epsfig{file=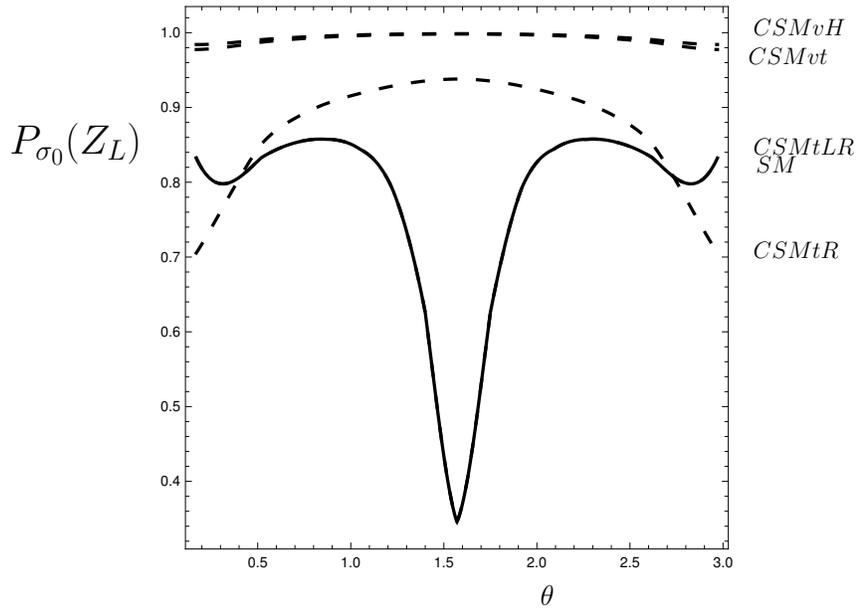, height=8.cm}
\]\\
\caption[1] {Energy dependence and angular distribution of $Z_L$ rate of $\sigma_0$ in SM, CSMtLR, CSMtR, CSMvt, CSMvH cases.}
\end{figure}

\clearpage

\end{document}